\documentclass[12]{article}
\usepackage{dcolumn}
\usepackage{color}
\usepackage{latexsym}
\usepackage{bm}
\begin{document}
\date{}

\title{Aspects of Diffeomorphism and Conformal invariance in 
classical Liouville theory}

\author{{Rabin Banerjee}$^{a,b}$\thanks{e-mail: rabin@bose.res.in}, \  
{Sunandan Gangopadhyay}$^{c,d}$\thanks{e-mail: sunandan.gangopadhyay@gmail.com, sunandan@bose.res.in}, \ 
{Shailesh Kulkarni}$^{a,e}$\thanks{e-mail: shailesh@bose.res.in}\\
$^a$\textit{S.N. Bose National Centre for Basic Sciences,}\\
\textit{JD Block, Sector III, Salt Lake, Kolkata-700098, India}\\
$^b$  \textit{Department of Physics, University of Helsinki}\\
\textit{ and Helsinki Institute of Physics,P.O.Box 64, FIN-00014 Helsinki, 
Finland}\\
$^c$ \textit{Department of Physics and Astrophysics, 
West Bengal State University, Barasat, India}\\
$^d$ \textit{Visiting Associate in 
S.N. Bose National Centre for Basic Sciences, Kolkata}\\
$^e$ \textit{Harish-Chandra Research Institute, Chhatnag Road, Jhusi,
Allahabad 211019, India}}

\maketitle
\begin{quotation}
\noindent \normalsize
The interplay between the diffeomorphism and conformal symmetries
(a feature common in quantum field theories)
is shown to be exhibited for the case of black holes 
in two dimensional classical Liouville theory. We show that although 
the theory is conformally invariant in the near horizon limit,
there is a breaking of the diffeomorphism symmetry at the classical level. 
On the other hand, in the region away from the horizon,
the conformal symmetry of the theory gets broken with the diffeomorphism symmetry remaining intact.
\end{quotation}

\noindent It is a well known fact from general
quantum field theoretic considerations \cite{polyakov}
that in two dimensions there is an interplay between
the diffeomorphism and the trace anomaly of the energy momentum tensor 
in the sense that it is possible
to remove one of them but not both. This is because for $D=2$ case, there
is no regularization which simultaneously preserves conformal as well as
diffeomorphism symmetries \cite{polyakov,green,bertlman,witten}.

\noindent Interestingly, we find that such a situation arises
in a classical Liouville theory \cite{liou} used to compute the black hole
entropy owing to the presence of a nontrivial central charge
in the theory \cite{solodukhin, cvitan,kang, pinamonti,pallua,chen}. Further, Hawking effect was also studied by using boundary Liouville model \cite{solodukhin2}.
In classical Liouville theory (which is relevant for black hole
physics) the energy momentum tensor
is covariantly conserved but has a nontrivial trace anomaly.
However, it turns out that if we consider terms upto leading order in the expansion of the metric, near the horizon, 
the trace anomaly vanishes and hence the theory becomes conformally invariant (which explains its
 utility for studies on black hole physics), but the energy-momentum tensor fails to remain covariantly conserved. Hence, the interplay between the diffeomorphism and the trace anomaly of the energy momentum tensor 
gets exhibited at the classical level.  

\noindent To begin with, we consider a four-dimensional 
spherically symmetric metric of the form
\begin{eqnarray}
ds^2 = \gamma_{ab}(x^0, x^1) dx^{a} dx^{b} 
+ r^{2}(x^0, x^1) d\Omega^{2} \label{0}
\end{eqnarray}
where $\gamma_{ab}(x^0, x^1)$ is the metric on an effective 
$2$-d spacetime $M^2$ with coordinates $(x^{0}, x^{1})$.
We now start from the four-dimensional Einstein-Hilbert action defined by
\begin{equation}
S_{EH} = -\frac{1}{16\pi G} \int_{M^4} \ d^{4}x 
\sqrt{-g} R_{(4)} \label{0.1}.
\end{equation}
Considering the above action on the class of 
spherically symmetric metrics (\ref{0}), 
we obtain an effective two-dimensional theory described by the action
\begin{equation}
S = - \int_{M^2} d^{2}x \sqrt{-\gamma} 
\left( \frac{1}{2}(\nabla\Phi)^2 + 
\frac{1}{4}\Phi^2 R_{(2)} + \frac{1}{2G}\right) \label{0.2}
\end{equation}
where $\Phi = rG^{-1/2}$ and 
$R_{(2)}$ is the two-dimensional scalar curvature
\footnote{From now onwards we shall suppress the suffix
on $R$ but it is understood as $2D$ curvature scalar.}. 
This action represents dilaton gravity in two 
dimensions with $r$ playing the role 
of dilaton field.

\noindent The above action can now be transformed 
to a form similar to that of the 
Liouville theory \cite{russo} 
\begin{equation}
S_{L} = - \int_{M^2} d^{2}x \sqrt{-\bar\gamma} \left(\frac{1}{2}
(\bar\nabla \phi)^2 + \frac{1}{4}q\Phi_{h}\phi \bar R
 + U(\phi)\right) \label{0.4}
\end{equation}  
with the aid of the following transformations 
\begin{equation}
\gamma_{ab} = \left(\frac{\phi_{h}}{\phi}
\right)^{\frac{1}{2}}e^{\frac{2\phi}{q\phi_{h}}}
\bar\gamma_{ab}\quad,  \   
\phi = \frac{\Phi^{2}}{q\Phi_{h}} \label{0.3} 
\end{equation}
where $\Phi_{h}=r_{h}G^{-1/2}$ 
is the classical value of the field $\Phi$ at the horizon and 
$U(\phi)=\frac{1}{2G}
\left(\frac{\phi_{h}}{\phi}\right)^{1/2}e^{\frac{2\phi}{q\phi_{h}}}$.
In (\ref{0.4}), $\bar\nabla$ is the 
covariant derivative compatible with the metric $\bar\gamma_{ab}$
while $\bar R$ is the corresponding curvature scalar.
\noindent Varying the above action with respect to 
the scalar field $\phi$ yields the equation of motion for $\phi$
\begin{equation}
\bar \nabla^{a}\bar \nabla_{a} \phi = \frac{q\Phi_{h}}{4} 
\bar R + \frac{dU}{d\phi}~. \label{new1}
\end{equation}
Similarly, varying the action with respect
to the metric $\bar\gamma_{ab}$, we obtain the constraints
\begin{equation}
T_{ab}\equiv {1\over 2} \partial_a\phi\partial_b
\phi-{1\over 4}\bar{\gamma}_{ab}(\bar\nabla\phi )^2
+{1\over 4}q\Phi_h (\bar{\gamma}_{ab}\bar \nabla^{c}\bar \nabla_{c} \phi 
-\bar\nabla_a\bar\nabla_b\phi)-{1\over 2}\bar{\gamma}_{ab}U(\phi)=0~.
\label{constra} 
\end{equation}
The theory of the scalar field $\phi$ 
described by the action
(\ref{0.4}) is not conformal in general. 
This can be easily seen by contracting (\ref{constra}) with
metric $\bar\gamma^{ab}$ to obtain 
\begin{equation}
 {T^{a}}_{ a} = \frac{q\Phi_{h}}{4} 
\bar \nabla^{a}\bar \nabla_{a} \phi - U(\phi) \label{new2}.
\end{equation}
By substituting (\ref{new1}) in the above equation 
we obtain on shell expression for
the trace of the energy-momentum tensor 
\begin{equation}
 {T^{a}}_{ a} = \left(\frac{q\Phi_{h}}{4}\right)^2 \bar R + \frac{q\Phi_{h}}{4}\frac{dU}{d\phi} - U(\phi)~.
\label{new2'}
\end{equation}
Thus, we have a nonvanishing trace of the energy-momentum tensor, 
leading to the breaking of the conformal symmetry 
in the classical Liouville theory. 
We would like to point out that 
such a violation in the conformal invariance generally occurs
when we quantize the theory on the curved background. 

\noindent We now take the divergence of (\ref{constra}) and obtain 
\begin{equation}
\bar\nabla_{a}{T^{a}}_{b} = 
\frac{1}{2}\left[\bar \nabla^{c}\bar \nabla_{c}\phi \bar\gamma_{be} 
-\frac{q\Phi_{h}}{4}\bar R \bar\gamma_{be}\right]
\bar\nabla^{e}\phi - \frac{1}{2}\partial_{b}U(\phi). \label{new3}
\end{equation}
Substituting the equations of motion for $\phi$ 
(\ref{new1}) in (\ref{new3}), leads to the
conservation of energy momentum tensor
\begin{equation}
\bar\nabla_{a}{T^{a}}_{b} = 0. \label{new4} 
\end{equation}
Hence, we find that although 
there is a  breaking of the classical conformal invariance as the
energy-momentum tensor has a nonvanishing trace (\ref{new2'}), 
the diffeomorphism symmetry remains intact. This is a feature
which has been observed earlier only in a quantum theory.

\noindent Now we consider the near horizon behavior of the theory 
defined by (\ref{0.4}). For simplicity
 we consider the metric
\begin{equation}
ds^2 = \bar{\gamma}_{ab}dx^{a}dx^{b}=
-g(x)dt^2 + \frac{1}{g(x)} dx^2  \label{new5}  
\end{equation}  
where $g(x =x_{h}) =0$ defines the location of the event horizon $x_{h}$ in Schwarzschild coordinates $(t,x)$. Since the metric coefficient $g(x)$ is a well behaved function, we can expand it about $x=x_{h}$
\begin{eqnarray}
g(x)=g'(x_h)(x-x_{h})+
\frac{1}{2!}(x-x_{h})^2 g''(x_h)+\frac{1}{3!}(x-x_{h})^3 g'''(x_h) + \cdots~.
\label{expan}
\end{eqnarray}
In the vicinity of the horizon, we keep terms proportional to $(x-x_{h})$ only. With this approximation,
the above function (\ref{expan}) becomes
\begin{equation}
 g(x) \approx g'(x_{h})(x-x_{h})=\frac{2}{\beta_{H}}(x-x_{h}) \label{new6}
\end{equation}
where $\beta_{H}$ is related to the Hawking temperature \cite{solodukhin}.
Now we would like to see how the metric function vanishes in the region near to the horizon.
For that it is convenient to introduce coordinate $z$ defined as
\begin{equation}
z= \int^x{dx\over g(x)}~.
\label{coord}  
\end{equation}  
Substituting (\ref{new6}) in (\ref{coord}) we get
\begin{equation}
g(x) \equiv g(z) 
= \frac{2}{\beta_{H}}e^{\frac{2z}{\beta_{H}}}.
\label{new7}
\end{equation}
Note that in the $(t, z)$ coordinates, 
large negative $z$ corresponds to near horizon limit.
We now consider the  equation of motion for scalar 
field (\ref{new1}) in the vicinity of the horizon.
Expressing (\ref{new1}) in $(t,z)$ coordinates we get
\begin{equation}
 -\partial_{t}^{2}\phi + \partial_{z}^{2}\phi = 
g(z)[\frac{q\Phi_{h}}{4}\bar R + U'(\phi)].\label{new8}
\end{equation}
In the region near to the horizon, in view of (\ref{new7}), 
the right hand side of above equation vanishes exponentially.
Hence, in the vicinity of the horizon we have
\begin{equation}
 -\partial_{t}^{2}\phi + \partial_{z}^{2}\phi = 0 .\label{new9}
\end{equation}
Therefore, in the near horizon region Liouville theory gets 
effectively described by free massless
scalar field. 
Noting that the trace of the energy-momentum tensor (\ref{new2})
can be written in the form (dropping the interaction term
$U(\phi)$ as it is not important for our present purpose)
\begin{equation}
{T^{a}}_{a}=\frac{1}{g(x)}(-T_{00}+T_{zz})
=-\frac{q\Phi_{h}}{4}\left[\frac{1}{g(x)}(\partial^{2}_{t}
\phi-\partial^{2}_{z}
\phi)\right]
\label{new10az}
\end{equation}
we find that
in the near horizon limit, using the near horizon 
equation of motion (\ref{new9}),
the trace of the energy-momentum
tensor as defined in \cite{solodukhin} becomes
\begin{equation}
-T_{00} + T_{zz} = 0. \label{new10}
\end{equation}
This  indicates the fact that near horizon theory is conformally invariant.
However, one can rewrite the right hand side of (\ref{new10az})
in terms of $\bar{R}$ using (\ref{new8}). This simply reproduces
the exact result for the trace of the energy-momentum tensor 
(\ref{new2'}). The near horizon result is trivially
obtained to be
\begin{equation}
{T^{a}}_{a}=-\left(\frac{q\Phi_{h}}{4}\right)^{2}g''(x_h).
\label{new10z}
\end{equation}
This shows that the above way of computing the trace near the 
horizon leads to a nonvanishing result 
which is incompatible with (\ref{new10}). To make the trace 
${T^{a}}_{a}$ compatible with (\ref{new10}), we shall use the equation
of motion (\ref{new9}) describing free massless scalar field
as the theory near the horizon has conformal invariance. This leads
to ${T^{a}}_{a}=0$ since the term in the braces in (\ref{new10az}) vanishes
as a consequence of (\ref{new9}). Hence, in the near horizon limit,
we take the equation of motion to be (\ref{new9}) with $g(x)$ vanishingly
small but not zero. We shall consistently apply this 
approximation in the subsequent near horizon analysis.

\noindent Now we show that, though we are able to keep the
conformal invariance intact, the near horizon theory does not preserve 
the diffeomorphism invariance. For that, we write the 
energy-momentum tensor (\ref{constra}) 
in the  $(t,z)$ coordinates as
\begin{eqnarray}
T_{00}(z) &=& \frac{1}{4}\left(\dot\phi^2 + 
(\partial_{z}\phi)^2 \right)
- \frac{q\Phi_{h}}{4}\left( \partial^2_{z}
\phi-\frac{g'(x)}{2}\partial_{z}\phi\right) 
\label{new15}\\
T_{0z}(z) &=& \frac{1}{2}\dot\phi~\partial_{z}\phi - 
\frac{q\Phi_{h}}{4} \left(\partial_{z}\dot\phi - 
\frac{g'(x)}{2}\dot\phi\right)\label{new16}\\
T_{zz}(z) &=& \frac{1}{4}\left(\dot\phi^2 + 
(\partial_{z}\phi)^2 \right)
 + \frac{q\Phi_{h}}{4}\left(-\ddot\phi + 
\frac{g'(x)}{2}\partial_{z}\phi\right) \label{new17} 
\end{eqnarray}
where the `dot' represents derivative 
with respect to time and `prime' represents derivative 
with respect to $x$. Note that once again we do not 
consider the interaction term $U(\phi)$ since
it is not important for our purpose. 
\noindent Now, let us consider the near horizon approximation 
of the above equations. 
First, we note that $T_{00}$, $T_{0z}$ and $T_{zz}$ 
contain a term proportional to a derivative of
the metric function. Therefore, it will be inappropriate 
to substitute the form of the metric function 
to first order in $(x-x_h)$ (see (\ref{new6})) in (\ref{new15}-\ref{new17}), 
rather, we have to take into account the next order term in the 
metric expansion (\ref{expan}). In other words, we put 
\begin{equation} 
g'(x) = \frac{2}{\beta_{H}} + g''(x_{h})(x-x_{h}) \label{new21}
\end{equation}
in the expressions for $T_{00}$, $T_{0z}$ and $T_{zz}$. Then we have
\begin{eqnarray}
T_{00}(z) &=& \frac{1}{4}\left(\dot\phi^2 + 
(\partial_{z}\phi)^2 \right)
- \frac{q\Phi_{h}}{4}\left( \partial^2_{z}
\phi-\frac{1}{2}[\frac{2}{\beta_{H}}+g''(x_{h})(x-x_{h})]
\partial_{z}\phi\right) 
\label{new29}\\
T_{0z}(z) &=& \frac{1}{2}\dot\phi~\partial_{z}\phi - 
\frac{q\Phi_{h}}{4} \left(\partial_{z}\dot\phi - 
\frac{1}{2}[\frac{2}{\beta_{H}}+g''(x_{h})(x-x_{h})]
\dot\phi\right)\label{new30}\\
T_{zz}(z) &=& \frac{1}{4}\left(\dot\phi^2 + 
(\partial_{z}\phi)^2 \right)
 + \frac{q\Phi_{h}}{4}\left(-\ddot\phi + 
\frac{1}{2}[\frac{2}{\beta_{H}}+g''(x_{h})(x-x_{h})]
\partial_{z}\phi\right) \label{new31}~. 
\end{eqnarray}
Further justification for this approximation 
will become clear in the subsequent discussion. 

\noindent We move on to compute the covariant divergence of
the energy-momentum tensor (\ref{new29}, \ref{new30}, \ref{new31}).
This can be written as :
\begin{eqnarray}
\bar\nabla_{a}{T^{a}}_{b} &=& \Lambda^{ac}\bar\nabla_{a}T_{cb}
\label{new32a+}
\end{eqnarray}
where, $\Lambda_{ab}$ is the metric in $(t, z)$ coordinates\footnote{Note 
that in the $(t, z)$ coordinates, 
the metric (\ref{new5}) reads $ds^2=-g(x)dt^2 + g(x)dz^2$. Hence, 
$\Lambda_{tt}=-g(x)$ and $\Lambda_{zz}=g(x)$.}.
For $b=t$, we obtain :
\begin{eqnarray}
\bar\nabla_{a}{T^{a}}_{t} &=& -\frac{1}{g(x)}
\left[\frac{1}{2}\dot{\phi}(\ddot{\phi}-\partial_{z}^{2}\phi)
-\frac{q\Phi_{h}}{8}g(x)g''(x_h)\dot{\phi}\right]~.
\label{new32b+}
\end{eqnarray}
It is worth mentioning now that rewriting the first term 
in (\ref{new32b+}) in terms
of $\bar{R}$ using the exact equation (\ref{new8}) 
yields a structure that precisely cancells the second term in (\ref{new32b+})
thereby leading to a vanishing covariant divergence 
of the energy-momentum tensor. This way, however, the effects of the
near horizon approximation are bypassed and the exact result is expectedly
reproduced. This is just the analogue of computing the trace (\ref{new10z})
by using the exact equation of motion (\ref{new8}). In order to
systematically implement the near horizon approximation 
that would be consistent with getting a vanishing trace (\ref{new10})
\cite{solodukhin}, we adopt the previous interpretation, that is take the
equation of motion as (\ref{new9}) with $g(x)$ 
vanishingly small but not zero. This leads to 
\begin{eqnarray}
\bar\nabla_{a}{T^{a}}_{t} &=& 
\frac{q\Phi_{h}}{8}g''(x_h)\dot{\phi}~.
\label{new32c+}
\end{eqnarray}

\noindent For $b=z$, we obtain 
(after using the near horizon equation of motion (\ref{new9})):
\begin{eqnarray}
\bar\nabla_{a}{T^{a}}_{z} &=& 
\frac{q\Phi_{h}}{8}g''(x_h)\partial_{z}\phi~.
\label{new32d+}
\end{eqnarray}

\noindent The above equations can be compactly written as :
\begin{eqnarray}
\bar\nabla_{a}T^{a}_{\quad b} &=& 
\frac{q\Phi_{h}}{8}g''(x_{h})\partial_{b}\phi~. \label{new32}
\end{eqnarray}
It is important to observe that the near horizon
equation of motion for $\phi$ (\ref{new9}) (which is different from the
equation of motion satisfied by $\phi$ away from the horizon (\ref{new1}))
plays an important role in the derivation of (\ref{new32}).
We therefore conclude that classical Liouville theory, when considered 
in the region near to the horizon, respects
conformal symmetry but it does not preserve the diffeomorphism invariance. 

\noindent This is a new result in our paper. We find that
though we are able to keep the conformal invariance intact, the near horizon
theory does not preserve the diffeomorphism invariance thereby 
clearly pointing out the interplay
between the diffeomorphism and conformal invariance exhibited
in the case of black holes in two dimensional classical Liouville theory.
To put this result in a proper perspective, we recall the findings of 
\cite{solodukhin} where it was shown that the conformal
symmetry near the horizon of the black hole leads to a Virasoro
algebra among the Fourier transform of specific combinations of 
the components of the energy-momentum
tensor. Here, we show that the Virasoro algebra which is a reflection
of conformal symmetry near the horizon of the black hole can also be
understood as the breaking of diffeomorphism symmetry near the black hole
horizon.

\noindent This observation is similar
to that of quantum anomalous theories. 
Indeed, when we quantize the scalar field theory on general curved
background, the trace of $\langle T_{ab}\rangle$ turns out to be nonzero. 
In particular, for 
the nonchiral theory in $1+1$ dimensions, 
$\langle {T^{a}}_{a}\rangle$ is proportional to the curvature scalar. 
However, the regularization adopted to compute the 
trace anomaly preserves the diffeomorphism invariance. One 
may adopt different type of regularization which spoils the diffeomorphism symmetry but keeps conformal symmetry intact. It turns out that in $D=2$ dimensions there is no regularization prescription 
which preserves the conformal as well as 
diffeomorphism invariance simultaneously \cite{polyakov,green}. 

\noindent As a side remark we mention that it might be possible to write an 
improved stress tensor from (\ref{new32}) that is conserved,
\begin{equation}
 \hat T^{a}_{\quad b} = T^{a}_{\quad b} - \frac{q\Phi_{h}}{8}g''(x_{h})\delta^{a}_{b}\phi~. \label{improvedEM}
\end{equation}
However, as is easily observed, this tensor is no longer traceless since,
\begin{equation}
 \hat T^{a}_{\quad a} = \frac{q\Phi_{h}}{4}\bar R \phi~. \label{improveEMtrace}
\end{equation}
Consequently, simultaneous imposition of both Ward identities is not
feasible. This shift of anomaly by using counterterms is more akin to what is
done in quantum field theory. A similar phenomenon for the Liouville theory
was also observed, though in a different context \cite{jackiw}.

\noindent The fact that in the region near the horizon, the energy-momentum tensor is not
covariantly conserved (\ref{new32}) can also be inferred by computing the classical Poisson algebra among the various light cone components of the energy-momentum tensor. It turns out that the Poisson algebra 
do not close which in general, is related to the breaking of either diffeomorphism or conformal invariance. 
This in fact justifies our approximation of keeping terms up to first order in $(x-x_h)$ in $g'(x)$
(\ref{new21}) in the evaluation of the covariant divergence of the energy-momentum tensor. It is 
important to note that if we compute $g'(x)$ from the near horizon expansion of $g(x)$ (\ref{new6}), it
would finally lead to the conservation of the energy-momentum tensor. This 
would be in direct clash with the non-closure of the Poisson algebra.

\noindent We begin our analysis of the classical Poisson algebra by writing (\ref{new29}-\ref{new31}) in the light cone coordinates 
\begin{eqnarray}
x^{+}&=&\frac{1}{\sqrt{2}}(t+z) \nonumber\\
x^{-}&=& \frac{1}{\sqrt{2}}(t-z)~. 
\end{eqnarray}
Then we have 
\begin{eqnarray}
T_{++} &=& T_{00} + T_{0z} \label{1.15}\nonumber\\
&=& \frac{1}{4} \left( \dot\phi + \partial_{z}\phi\right)^2 
 - \frac{q\Phi_{h}}{4}\left( \partial^{2}_{z}\phi + \partial_{z}\dot\phi
 -\frac{1}{2}[\frac{2}{\beta_{H}}+\frac{g''(x_{h})}{2}(x-x_{h})](\partial_{z}\phi + \dot\phi)\right) \label{new18}\\
T_{--} &=& T_{00} - T_{0z}\label{1.25}\nonumber\\
&=& \frac{1}{4} 
\left( \dot\phi - \partial_{z}\phi\right)^2 
 - \frac{q\Phi_{h}}{4}
\left( \partial^{2}_{z}\phi - \partial_{z}\dot\phi  
 - \frac{1}{2}[\frac{2}{\beta_{H}}+\frac{g''(x_{h})}{2}(x-x_{h})](\partial_{z}\phi - \dot\phi)\right) ~. \label{new20}
\end{eqnarray}
Next, we give the basic Poisson brackets among the canonical variables 
\cite{solodukhin} 
\begin{eqnarray}
 \{\phi(z),\dot\phi(z')\} &=& \delta(z-z') \label{new22}\\
 \{\phi(z),\phi(z')\} &=& 0\label{new23}\\
  \{\dot\phi(z),\dot\phi(z')\} &=& 0 \label{new24}~.
\end{eqnarray}
The Poisson algebra between $T_{++}(z)$ and $T_{++}(z')$ therefore reads 
\begin{eqnarray}
\{T_{++}(z), T_{++}(z')\} &=& 
\left(T_{++}(z) + 
T_{++}(z')\right)\partial_{z}\delta(z-z') \nonumber\\
&& + \frac{q^2\Phi_h^{2}}{8}\left[ -\partial_{z}^3 
\delta(z-z') + \frac{g'(x)g'(x')}{4}
\partial_{z}\delta(z-z') + \frac{1}{2}[g'(x)-g'(x')]
\partial_{z}^2 \delta(z-z')\right]
\label{new25}
\end{eqnarray}
 with $g'(x)$ given by (\ref{new21}). Note that here, we have only given the Poisson bracket among  $T_{++}$ component of EM tensor because the algebra among $T_{--}(z)$ and 
$T_{--}(z')$  is identical in structure, while the Poisson bracket among $T_{++}$ and $T_{--}$ is zero. 
Now neglecting terms proportional to $(x-x_{h})$ in the above algebra leads to 
\begin{eqnarray}
\{T_{++}(z), T_{++}(z')\} &=& \left(T_{++}(z) + 
T_{++}(z')\right)\partial_{z}\delta(z-z') \nonumber\\
&&+\frac{q^2\Phi_h^{2}}{8}\left[ -\partial_{z}^3 
\delta(z-z') + \frac{1}{\beta_{h}^2}\partial_{z}\delta(z-z')\right]~.\label{new26}
\end{eqnarray}
This algebra would have been the same if we had substituted $g'(x) = \frac{2}{\beta_{H}}$ in 
(\ref{new15}-\ref{new17}). However, in that case we would not have found any violation in the conservation
of the energy-momentum tensor (\ref{new32}) which would be inconsistent with the fact that the Poisson algebra (\ref{new26}) does not close.    

\noindent  For the sake of completeness, we now move on to investigate 
the Virasoro algebra. The Virasoro generator 
is defined as \cite{solodukhin}
\begin{equation}
L_{n} = \frac{L}{2\pi}  \int_ {L/2}^{L/2} 
 dz~ T_{++}(z) e^{2i\pi nz/L}~.  \label{new27}
\end{equation}

\noindent From the algebra given in (\ref{new26}), we compute the
algebra between the Virasoro generators which yields :
\begin{eqnarray}
\{L_n, L_m\}&=&(n-m)L_{n+m} + 
\frac{c}{12}n\left(n^2+\left(\frac{L}{2\pi\beta_h}\right)^2\right) 
\delta_{n+m,0}~.
\label{vira_alg}
\end{eqnarray} 
This is the expression for the classical Virasoro algebra 
with the central charge $c=3\pi q^2\Phi^{2}_{h}$
\cite{solodukhin}.  \\

\noindent Discussions :\\

\noindent In this article we have studied thoroughly the interplay 
between
the diffeomorphism and conformal symmetries for black holes 
in $D=2$ classical Liouville
theory. The energy-momentum tensor derived from $D=2$ classical 
Liouville action, in general, is covariantly conserved. 
However, the trace of the energy-momentum tensor 
becomes nonzero leading to violation of the conformal invariance.

\noindent In the region near to the horizon, Liouville theory shows 
interesting behavior. In the vicinity of the horizon we expand 
the metric function about the horizon and keep only the leading order
terms. In this approximation, the equations of motion for Liouville field
takes the form of $D=2$ free, massless Klein-Gordon equation and eventually
makes the energy-momentum tensor traceless. Hence, in the 
vicinity of the horizon
Liouville theory is conformally invariant. However, we observe that, in 
the near horizon limit the energy-momentum tensor is not covariantly conserved. In fact, 
we showed that the covariant divergence of energy-momentum 
tensor is proportional
to the curvature scalar. This fact is quite well known in the context of quantization
of fields on the general curved background. Classically, the energy-momentum tensor, of a field theory
under consideration, is traceless and also covariantly conserved. However, during the process
of quantization, either there is a trace or diffeomorphism anomaly, depending upon
the choice of regularization. 
It is impossible to preserve both, the conformal as well as
the general coordinate  (diffeomorphism) 
invariance \cite{polyakov, green}. In this
paper, for the classical Liouville theory, 
we have a similar kind of behavior of the energy-momentum
tensor. Away from the horizon, the classical energy-momentum
tensor is covariantly conserved signalling the presence of diffeomorphism
invariance but the conformal symmetry is lost due to a nonvanishing trace.
Near the horizon, on the contrary, the diffeomorphism symmetry
is broken but the conformal symmetry is intact. This interplay is a new finding
in the context of classical Liouville 
theory in the near horizon approximation.

\noindent Another feature in our paper is that the Virasoro
algebra which is a reflection of the conformal invariance near the
horizon of the black hole can also be understood as the breaking of the
diffeomorphism invariance near the black hole horizon.

\noindent Here we would like to mention that similar results, 
though in a different 
context, were discussed in \cite{jackiw}. 
To be more specific counterterms were added 
in the parent $D=2$ Liouville action in order 
to restore Weyl invariance which however,
breaks diffeomorphism symmetry. 
The point is that counterterms, taken in \cite{jackiw},
are a manifestation of regularization ambiguities. 
However, the issue of regularization
is meaningful only in the quantum field theory. 
In our example no counterterms are
necessary. The interplay of the symmetries occurs very naturally 
with the near horizon
results displaying one feature while the away from the horizon displaying another
feature. The whole analysis, either at the technical or conceptual level, is completely
classical. \\

\noindent {\bf Acknowledgements}\\

\noindent One of the authors (RB) thanks Masud Chaichian and Anca
Tureanu for their kind hospitality at the Physics Department, University of
Helsinki, where part of this work was done.
The authors would also like to thank the referee for useful comments.



\end{document}